# Quantum dots on the InAs(110) cleavage surface created by atom manipulation

Van Dong Pham[1,*], Yi Pan[1,2,*], Steven C. Erwin[3], and Stefan Fölsch[1,#]

[1]*Paul-Drude-Institut für Festkörperelektronik, Hausvogteiplatz 5-7, Leibniz-Institut im Forschungsverbund Berlin e. V., 10117 Berlin, Germany*

[2]*Center for Spintronics and Quantum Systems, State Key Laboratory for Mechanical Behavior of Materials, Xi'an Jiaotong University, Xi'an 710049, China*

[3] *Center for Computational Materials Science, Naval Research Laboratory, Washington, DC 20375, USA*

**Abstract**

Cryogenic scanning tunneling microscopy was employed in combination with density-functional theory calculations to explore quantum dots made of In adatoms on the InAs(110) surface. Each adatom adsorbs at a surface site coordinated by one cation and two anions, and transfers one electron to the substrate, creating an attractive quantum well for electrons in surface states. We used the scanning-probe tip to assemble the positively charged adatoms into precisely defined quantum dots exhibiting a bound state roughly 0.1 eV below the Fermi level at an intrinsic linewidth of only ~4 meV, as revealed by scanning tunneling spectroscopy. For quantum-dot dimers, we observed the emergence of a bonding and an antibonding state with even and odd wavefunction character, respectively, demonstrating the capability to engineer quasi-molecular electronic states. InAs(110) constitutes a promising platform in this respect because highly perfect surfaces can be readily prepared by cleavage and charged adatoms can be generated *in-situ* by the scanning-probe tip.

**Orcid IDs**

VDP: 0000-0002-1416-2575

YP: 0000-0003-1978-475X

SCE: 0000-0002-9675-9411

SF: 0000-0002-3336-2644

[*] VDP and YP contributed equally to the work

[#] foelsch@pdi-berlin.de

Semiconductor quantum dots play a central role in optoelectronic device applications [1]. At the level of fundamental research, they make it possible to explore superimposed and entangled quantum states [2,3], semiconductor qubits [4], electron correlation in artificial lattices [5], and electronic quantum transport [6], to mention only a few. Aside from their fabrication in the form of colloidal crystals [7], quantum dots are typically created in semiconductor heterostructures by growing vertically and laterally aligned nanocrystals [2,3,8,9], by imposing lateral confinement using electron-beam lithography [5,10], or by depleting a two-dimensional electron gas (2DEG) using external gates [11,12] and local oxidation [13].

The method of 2DEG depletion using external gates exploits the electric field effect to spatially modulate the carrier density. In our previous work [14], we followed a similar idea of spatially controlling the electrostatic surface potential, however, at the level of single atoms: we used the tip of a scanning tunneling microscope (STM) to assemble short atomic chains on an InAs(111)A surface by atom manipulation [15]. The chains consisted of six positively charged In adatoms leading to electronic confinement, and hence, the emergence of a bound state with discrete energy – the fingerprint of a quantum dot. We showed that these dots can be arranged and thereby coupled in various ways, yielding quasi-molecular electronic states which can be described by a tight-binding Hamiltonian assuming a single s orbital on each site [14,16].

Here, we extend this concept to the (110) cleavage surface of indium arsenide. First, the adsorption and charge state of an In adatom on InAs(110) are investigated by STM and complementary density-functional theory (DFT) calculations. It is then demonstrated that an assembly of six adatoms confines electrons and thus behaves like a quantum dot. The resulting bound state has an intrinsic linewidth of ~4 meV which is remarkably small for confined electronic states at surfaces. Finally, it is shown that quantum-dot dimers can be created, leading to bonding and antibonding states as verified by scanning tunneling spectroscopy (STS). The present findings are important because they facilitate the STM-based construction of quantum structures on a new InAs platform: a cleaved (110) surface is significantly easier to prepare than a (111)A-terminated surface requiring a dedicated growth facility for molecular beam epitaxy (MBE) [17]. Moreover, working in (110) surface orientation offers the prospect of exploring cleaved III-V semiconductor heterostructures in cross-sectional geometry to ultimately be able to create electrical gating of the STM-generated nanostructures.

The STM investigations were carried out in ultrahigh-vacuum (UHV) at a sample temperature of 5 K. We used undoped and (001)-oriented InAs wafers cleaved in UHV to obtain the InAs(110) surface. The left panel in Fig. 1(a) shows a constant-current topography image of an In adatom adsorbed on the InAs(110) surface. At the sample bias of 0.1 V as applied here, the surface As atoms are imaged as protrusions arranged in rows along the [1$\bar{1}$0] in-plane direction [18]; the spacing between the rows is $a_0$=6.06 Å, the cubic lattice constant of InAs. The adatom in the center of the image is located in the channel in between of the rows, consistent with previous work [19,20] predicting interstitial configurations in which the adatom is bonded either to two cations and one anion (labeled $I_{i1}$) or to two anions and one cation ($I_{i2}$). In agreement with the findings by Weber *et al.* [20], our DFT calculations confirm that $I_{i2}$ is the most stable configuration, as illustrated by the right hand-side panel showing a close-up view together with an overlaid structure model. (The orientation of the structure model with respect to the (001) plane is confirmed experimentally by the chemical contrast between surface anions and cations (Supplementary Material, Fig. S1) – first revealed by the atom-selective STM imaging of GaAs(110) [21].)

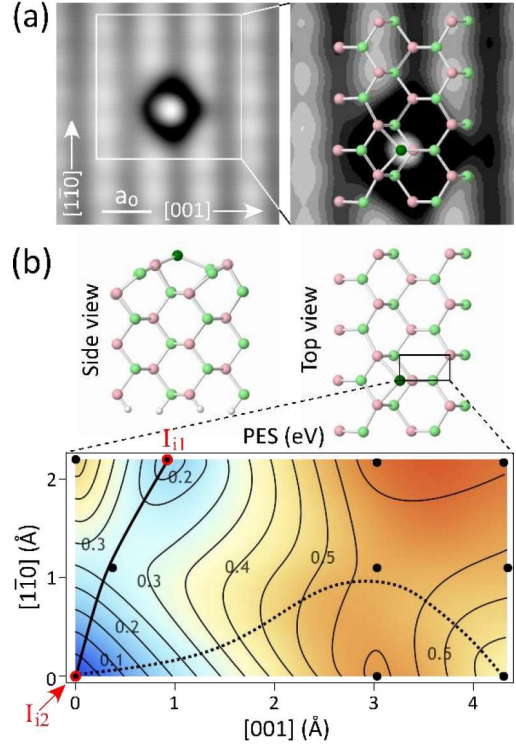

*Fig. 1. (a) Left: STM image (1 nA, -0.6 V) of an In adatom on InAs(110) located within the channel between rows of surface As atoms imaged as protrusions, row spacing: $a_0$=6.06 Å. Right: close-up view at staggered gray scale with overlaid structure model showing the surface In (green) and As atoms (pink) as well as the In adatom (dark green). (b) Top: structural model in top and side view as determined by DFT. Bottom: DFT potential-energy surface (PES) for an In adatom diffusing on InAs(110); the solid line shows the minimum energy path between the stable ($I_{i2}$) and metastable ($I_{i1}$) configurations along the channel whereas the dotted line indicates the barrier across the channel. At each black point within the rectangle, the x- and y-position of the adatom were held fixed while all other atomic coordinates (including the first four InAs layers) were relaxed.*

We performed DFT calculations to determine the equilibrium geometry of InAs(110) with and without adsorbed In adatoms, as well as the potential-energy surface for surface diffusion of those adatoms (general details on the calculations are given in Ref. [14]). Figure 1(b) shows the DFT potential-energy surface for an In adatom diffusing on the InAs(110) cleavage surface and provides quantitative information on the adsorption behavior: the metastable configuration $I_{i1}$ is 0.2 eV higher in energy than the stable configuration $I_{i2}$. The diffusion barrier along the channel is 0.3 eV (the minimum energy path is sketched as a solid curve); this barrier can be overcome above ~120 K. On the other hand, the barrier across the channel is about 0.6 eV (dotted curve) and can be overcome above ~240 K. However, this pathway is probably preempted by exchange between the adatom and a surface In atom, similar to the exchange reaction previously observed for Mn substitutional impurities on InAs(110) [22].

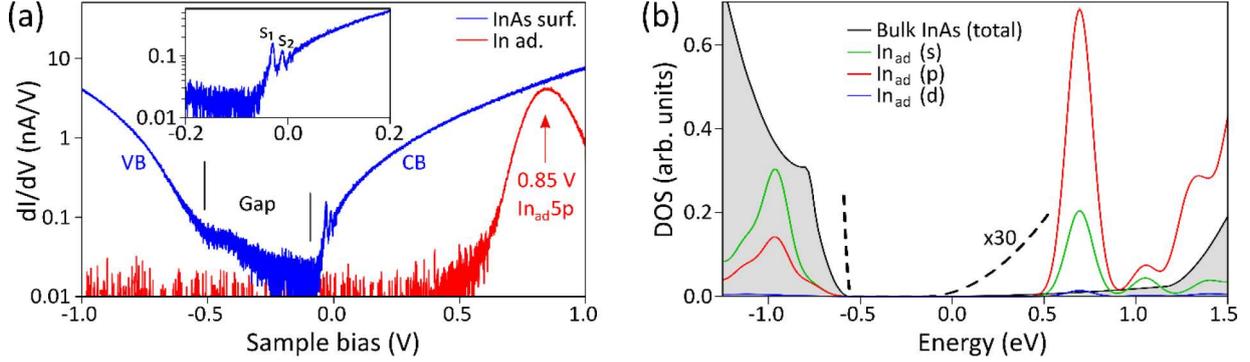

Fig. 2 (a) Conductance spectra on a logarithmic scale recorded with the tip probing the bare surface (blue) and a discrete In adatom (red). The blue spectrum indicates Fermi level pinning in the conduction band (CB) and the presence of accumulation-layer states denoted $s_1$ and $s_2$ near the CB minimum, see inset; the red spectrum reveals an adatom-derived unoccupied state at 0.85 eV. (b) Electronic structure of In on InAs(110) calculated by DFT with the Heyd-Scuseria-Ernzerhof screened hybrid functional; projected bulk bands are gray (scaled up to highlight the band onsets, see dashed line), the zero of energy was set 100 meV above the CB minimum. Colored curves show the density of states projected onto the s-, p-, and d- valence orbitals of the adatom, the experimental state at 0.85 eV derives predominantly from 5p atomic orbital states.

We probed the electronic surface properties by STS measurements of the tunnel conductance dI/dV which provides an approximate measure of the electronic density of states. The spectrum shown blue in Fig. 2(a) was recorded with the tip probing the bare surface. It reveals the energy band gap of InAs (0.42 eV at the measurement temperature of 5 K [23]) and, most prominently, Fermi-level pinning in the conduction band. (The residual conductance observed within the band gap is due to the so-called "dopant-induced" contribution arising from electrons tunneling out of the filled conduction-band states located near the band edge [24].) Fermi level pinning in the conduction band is a generic feature of InAs surfaces, indicating charge accumulation at the surface [20]. In agreement with previous STS work on cleaved InAs(110) [18], we observe conductance peaks near the conduction-band minimum just below the Fermi level (at sample bias V=0) which are a manifestation of conduction-band states that undergo vertical confinement – and thereby quantization – because of the downward band bending near the surface. In the blue spectrum in Fig. 2(a), these states show up as two peaks at -30 and -11 mV, respectively, reflecting the two lowest subbands of the accumulation layer (denoted $s_1$ and $s_2$). The actual energy and magnitude of the peaks observed depends on the location probed by the tip, see Fig. S2 in the Supplementary Material. We attribute this spatial variation to the effect of electrostatic disorder due to defects [25,26,27] in the surface-near region.

The red spectrum in Fig. 2(a) was recorded with the tip probing a single In adatom on InAs(110) and reveals a conductance peak at a sample bias of 0.85 V. The corresponding state derives predominantly from 5p atomic orbital states of the adatom as evident from the density of states calculated by DFT, see Fig. 2(b). It is noted that a similar adatom-induced state was observed previously for In adatoms on InAs(111)A [28] and GaSb(110) [29].

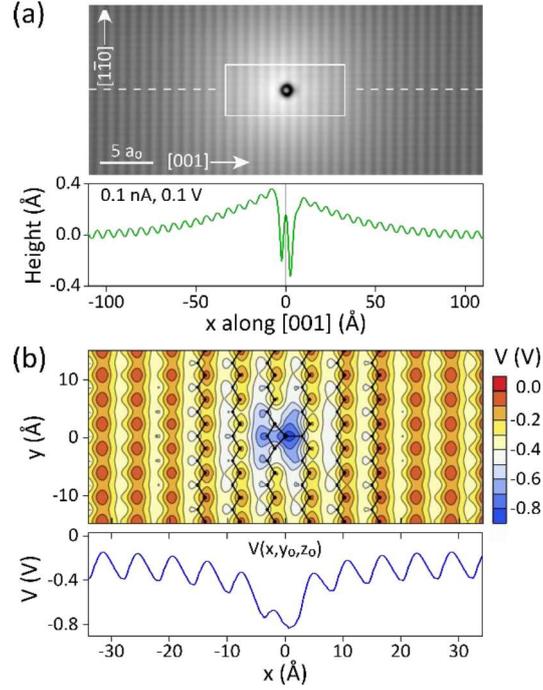

*Fig. 3. (a) Upper panel: larger-scale STM image (0.1 nA, 0.1 V) of an In adatom on InAs(110); the increased apparent height around the positively charged adatom arises from its screened Coulomb potential. Lower panel: measured height contour taken along the dashed line in (a) showing that the local band bending is asymmetric about the adatom position. (b) Upper panel: DFT electrostatic potential 3 Å above the surface inside the rectangular box in (a). Blue shading is more strongly attractive for electrons. The In adatom and the adjacent In surface atom together have a net positive charge +1e, explaining the experimentally observed asymmetry. Lower panel: line scan of the calculated potential $V(x,y_0,z_0)$ showing the two separate contributions from the adatom and the surface In atom.*

Similar to the situation observed on InAs(111)A, In adatoms on InAs(110) are positively charged. This is evident from the increased apparent height around the charged adatom when imaged at positive sample bias as in the upper panel of Fig. 3(a). The increased height is due to the screened Coulomb potential induced by the charged adatom which locally increases the density of states available for the tunnel process [30,31]. It is noteworthy that the hillock produced by the local potential is not symmetric about the adatom position. This asymmetry is clearly evident from the topographic line scan in the lower panel of Fig. 3(a) taken along the dashed line.

To explain this observation, we consider the DFT electrostatic potential in a plane just above the surface. A contour plot of the potential, 3 Å above the surface and within the rectangular area marked in the STM image, is shown in the upper panel of Fig. 3(b). Far from the adatom, the local potential reflects the fact that electrons are transferred from surface In atoms to surface As atoms in accordance with the electron counting rule. The defect complex consisting of the In adatom plus the surface In atom to its left transfers one electron to the environment, creating a localized positive charge that is attractive for electrons. The line scan in the lower panel of Fig. 3(b) shows the contributions from the adatom and the surface In atom. The double-atom structure of this defect complex explains the asymmetry observed in STM topography.

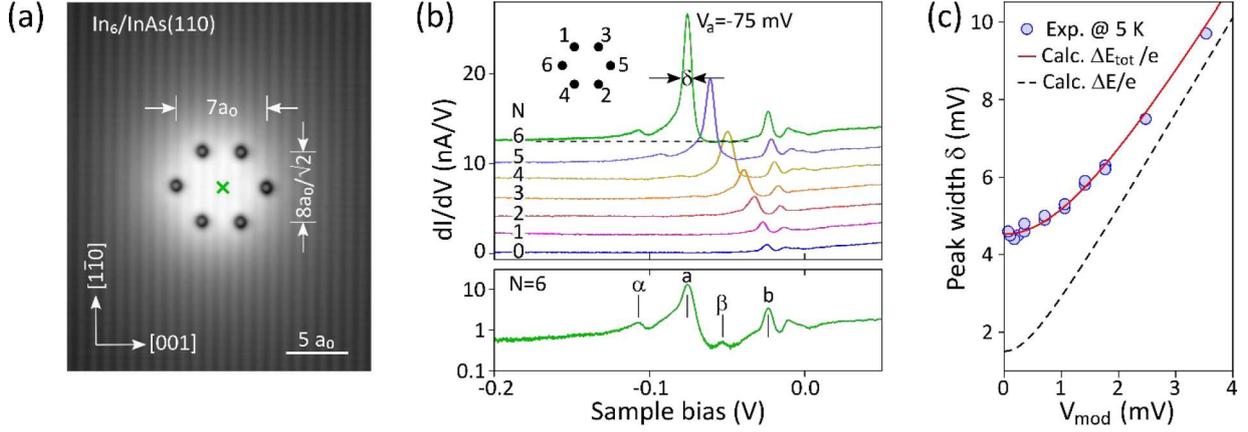

Fig. 4. (a) STM image (0.1 nA, 0.1 V) of six In adatoms on InAs(110) assembled into a hexagon ("dot") $7a_0$=42.42 Å wide and $8a_0/\sqrt{2}$=34.28 Å tall, the increased apparent height is due to the Coulomb potential arising from the adatoms. (b) Upper panel: size-dependent conductance spectra with N the number of adatoms in the dot (assembly sequence highlighted by the inset): a strong peak evolves at $V_a$=-75 mV for the complete hexagon (N=6) together with smaller peaks at higher and lower energies; the width $\delta$ of the main peak (full width at half maximum) is measured relative to the dashed horizontal baseline. Lower panel: logarithmic plot of the N=6 spectrum, peaks denoted $\alpha$ and $\beta$ are replicas of peaks a and b induced by inelastic electron tunneling. (c) Measured peak width $\delta$ (blue symbols) versus lock-in modulation voltage $V_{mod}$ (root-mean-square value); the dashed curve indicates the resolution limit $\Delta E/e$ while the red curve is the calculated peak width $\Delta E_{tot}/e$ assuming an intrinsic line width $\Gamma$=4.3 meV of the corresponding state.

In contrast to the MBE-grown InAs(111)A surface [32], cleaved InAs(110) is free of native adatoms. Nevertheless, individual In atoms can be transferred from the scanning-probe tip to the surface [33] after proper preconditioning of the tip [34] and subsequently repositioned by lateral atom manipulation (Supplementary Material, Fig. S3). This method was used in Fig. 4(a) to arrange N=6 adatoms into a hexagon having a width of $7a_0$=42.42 Å along [001] and $8a_0/\sqrt{2}$=34.28 Å along the [1$\bar{1}$0] direction. The upper panel of Fig. 4(b) shows a set of size-dependent conductance spectra intermediately recorded at the tip position marked in the STM image in Fig. 4(a) while building up the hexagon one adatom after another. Starting from the bare surface [N=0, same spectrum as in Fig. 2(a)], it is seen that a strong conductance peak evolves for the completed hexagon (N=6), revealing a bound state 75 meV below the Fermi level. Besides this dominant spectral feature, a number of smaller peaks are observed as readily seen in the logarithmic plot of the N=6 spectrum in the lower panel of Fig. 4(b).

These spectra demonstrate that the attractive potential induced by the assembled adatoms confines electrons at the cleaved InAs(110) surface. Hence, the hexagon acts as a quantum dot – an "artificial atom" – that creates a bound state of discrete energy. (It is noted that electron confinement was reported recently also for Cs adatom structures created by atom manipulation on the InSb(110) surface [35].) It is tempting to interpret the present spectra in the way that the dot gives rise to a lateral confinement of accumulation-layer states with the bound state at -75 meV deriving from the lowest subband. This alone, however, provides no conclusive picture why the lowest subband would be confined in the first place rather than higher-lying subbands. (All spectral features observed here were reproduced for various independently assembled dots in different experimental runs.) It remains to be established precisely which surface states are confined in the present case; the central finding of this work is that quantum dots can be assembled from charged In adatoms on the cleaved InAs(110) surface.

Returning to the logarithmic plot in Fig. 4(b), the conductance peaks labeled $\alpha$ and $\beta$ are correlated to those labeled $a$ and $b$: the difference in bias voltage at which they occur is $V_a - V_\alpha = 32$ mV and, similarly, $V_b - V_\beta = 30$ mV. This suggests that $\alpha$ and $\beta$ are replicas of $a$ and $b$, respectively, induced by inelastic electron tunneling. The corresponding energy transfer observed here is consistent with the excitation of optical phonons implying a transfer of 29 meV as found previously by high-resolution electron-energy loss spectroscopy on cleaved InAs(110) [36].

The bound state of the quantum dot – reflected by peak $a$ in Fig. 4(b) – has an exceptionally small line width as deduced from tunnel conductance measurements at varying energy resolution. We recorded corresponding peak profiles at different root-mean-square values $V_{mod}$ of the lock-in modulation voltage and found that the profile steadily sharpens as $V_{mod}$ is reduced. At successively small modulation, the measured peak width $\delta$ (the full width at half maximum [37]) converged to a value well below 5 mV as evident from the data points collected in Fig. 4(c). The theoretical energy resolution [38] is $\Delta E = \sqrt{(3.5kT)^2 + (2.5eV_{mod})^2}$, where the two terms in the square root take into account the broadening due to temperature and lock-in modulation, respectively. The black dashed line in Fig. 4(c) shows the quantity $\Delta E/e$ at the measurement temperature of 5 K. On the other hand, a state with intrinsic line width $\Gamma$ is expected to be observed at a total energy broadening of $\Delta E_{tot} = \sqrt{(3.5kT)^2 + (2.5eV_{mod})^2 + \Gamma^2}$. The quantity $\Delta E_{tot}/e$ defined by the latter expression yields an excellent fit of the data points at an intrinsic line width of $\Gamma = 4.3$ meV as shown by the full red curve in Fig. 4(c).

Next, we demonstrate that quasi-molecular electronic states can be created by bound-state coupling in quantum-dot dimers created on the InAs(110) cleavage surface. In the topography image displayed in Fig. 5(a), we started from the same dot as in Fig. 4(a) and added an identical dot at a center-to-center spacing of $17a_0/\sqrt{2} = 72.85$ Å along the [1$\bar{1}$0] direction, yielding a quantum-dot dimer. The corresponding spectra in Fig. 5(b) were recorded with the tip placed above (blue) and in between of the dots (red). They reveal the emergence of a bonding ($\sigma$) and an antibonding state ($\sigma*$) as expected for the symmetric and antisymmetric superposition of the bound states belonging to the two dots. The observed $\sigma$–$\sigma*$ splitting measures $\Delta_{\sigma-\sigma*} = 75$ mV. Note also that the $\sigma$–$\sigma*$ doublet is downshifted from the energy of the single dot indicated by the dashed line; this shift arises from the electrostatic potential change that each dot experiences from the other as previously observed in quantum-dot dimers on InAs(111)A [16]. Aside from the dominant $\sigma$ and $\sigma*$ peaks in the conductance spectra, we again observe a set of smaller peaks at higher and lower energies as detailed in the lower panel of Fig. 5(b). Again, the spectra feature inelastic replicas consistent with the excitation of surface phonons [36]; this is evident from the bias-voltage differences of 31, 33, and 30 mV found for the conductance-peak pairs ($\gamma,g$), ($\nu,n$), and ($\tau,t$), respectively. Finally, to further analyze the bonding and antibonding states, Fig. 5(c) displays spatial conductance maps recorded at the corresponding bias voltages of the σ and σ∗ peaks in panel (b), respectively. These maps are consistent with the symmetric (σ) and antisymmetric (σ∗) wave-function character of the bonding and antibonding states.

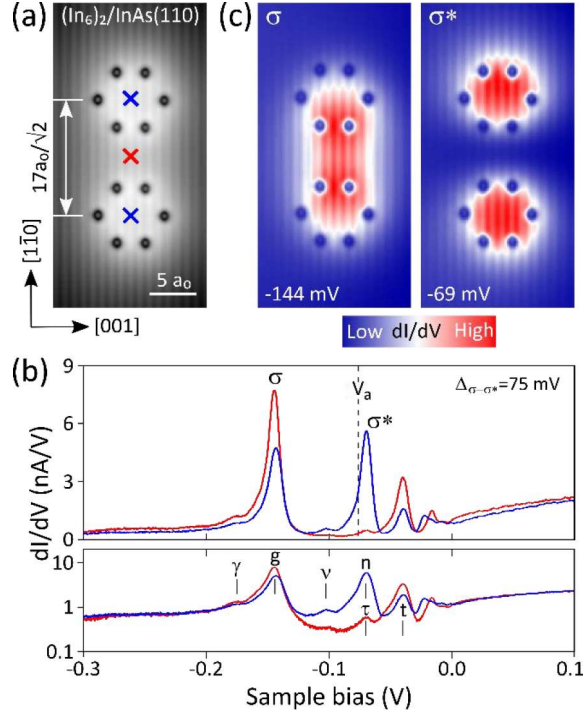

*Fig. 5. (a) STM image (0.1 nA, 0.1 V) of the same dot as in Fig. 4 (a) after adding another dot of identical structure at a center-to-center spacing of $17a_0/\sqrt{2}=72.85$ Å along $[1\bar{1}0]$. (b) Upper panel: spectra recorded with the tip placed above (blue, average of equivalent spectra taken on either dot) and in between of the dots (red): two major peaks labeled σ and σ∗ are observed together with smaller peaks; the vertical dashed line marks the peak position initially observed for the single dot. Lower panel: logarithmic plot of the same spectra featuring replicas of conductance peaks due to inelastic electron tunneling. (c) Spatial conductance maps recorded at the sample biases where the σ and σ∗ peaks are observed in (b), confirming the bonding (left) and antibonding character (right) of the corresponding states of the dimer.*

In conclusion, we employed cryogenic STM and STS measurements in combination with DFT calculations to perform an in-depth study of individual In adatoms on the InAs(110) cleavage surface. The adatom is found to adsorb in the channel between the rows of surface atoms along the $[1\bar{1}0]$ in-plane direction and is bonded to two anions and one cation ($I_{i2}$ configuration), in agreement with previous theoretical work [20]. The $I_{i1}$ configuration (bonding to two cations and one anion) is metastable and 0.2 eV higher in energy. It is found that the adatom together with its neighboring surface In atom along the $[00\bar{1}]$ in-plane direction are positively charged, with a net charge of +1e. By arranging these charged adatoms into groups with atomic precision, we have created quantum dots that laterally confine electrons in surface states. For the bound state resulting from this confinement we found an intrinsic line width of only ~4 meV. (The linewidths observed in conductance spectra of confined electrons on metal surfaces are typically larger by a factor of

10 to $10^2$ [39,40].) Finally, it was demonstrated that the quantum coupling between two identical dots placed side by side leads to the emergence of a bonding and an antibonding state indicating the symmetric and antisymmetric superposition of the dot wave functions.

The comparably small intrinsic line width observed for the bound state of atomic-scale quantum dots on InAs(110) will open the way to resolve the energy level spectrum of more elaborate quantum-dot arrays hosting exotic electronic states [41,42]. The system presented here has the practical advantage that large-scale and atomically flat surface terraces are readily prepared by cleavage in UHV and no extra deposition of material is required to generate charged adatoms, making it a promising platform for the creation of semiconductor quantum structures by scanning-probe techniques.

**ACKNOWLEDGEMENTS**

VDP, YP, and SF acknowledge funding by the Deutsche Forschungsgemeinschaft (DFG, German Research Foundation) – FO362/4-2, 437494632. YP also acknowledges the National Natural Science Foundation of China (12074302) for supporting the collaboration. SCE was supported by the Office of Naval Research through the Naval Research Laboratory's Basic Research Program. We gratefully acknowledge fruitful discussions with Klaus Biermann and Randall M. Feenstra.

# Supplementary Material to:

# Quantum dots on the InAs(110) cleavage surface created by atom manipulation

Van Dong Pham[1,*], Yi Pan[1,2,*], Steven C. Erwin[3], and Stefan Fölsch[1,#]

[1]*Paul-Drude-Institut für Festkörperelektronik, Hausvogteiplatz 5-7, Leibniz-Institut im Forschungsverbund Berlin e. V., 10117 Berlin, Germany*

[2]*Center for Spintronics and Quantum Systems, State Key Laboratory for Mechanical Behavior of Materials, Xi'an Jiaotong University, Xi'an 710049, China*

[3] *Center for Computational Materials Science, Naval Research Laboratory, Washington, DC 20375, USA*

**Contents:**

S1. Chemical contrast in bias-dependent surface imaging

S2. Accumulation-layer states on bare InAs(110)

S3. Vertical and lateral atom manipulation of In on InAs(110)

[*] VDP and YP contributed equally to the work

[#] foelsch@pdi-berlin.de

## S1. Chemical contrast in bias-dependent surface imaging

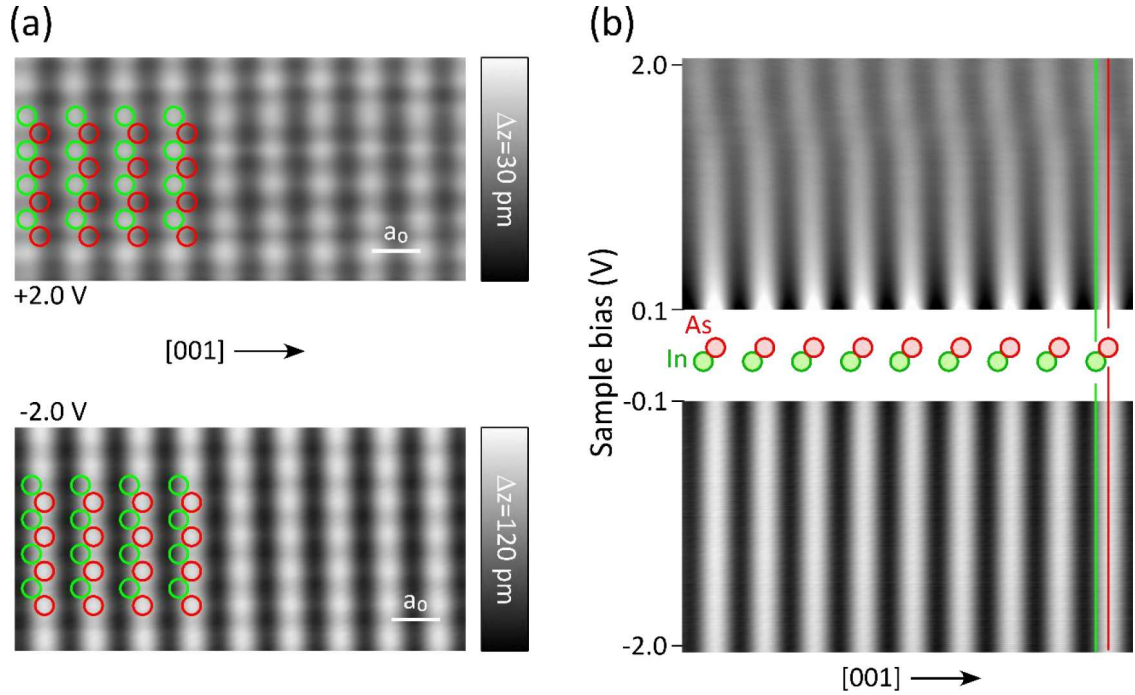

Fig. S1. (a) Constant-current topography images recorded at a tunnel current of 1 nA and sample biases of -2 V (black-to-white contrast Δz=120 pm, bottom) and +2 V (Δz=30 pm, top). Similar to the atom-selective STM imaging of GaAs(110) reported in the seminal work of Feenstra *et al.* (Ref. 22 in the main text), the occupied state density is concentrated around the surface anions (As) while the unoccupied state density is concentrated around the surface cations (In). Consequently, at -2 V the surface As atoms appear as protrusions (marked by red circles) while at +2 V the surface In atoms appear as protrusions (green circles). (b) Bias-dependent height maps showing constant-current height profiles recorded along the [001] in-plane direction perpendicular to the rows between -2.0 and -0.1 V (occupied states, lower panel) and between 0.1 and 2.0 V (unoccupied states, upper panel). The change in chemical contrast occurs at roughly 1.5 V.

## S2. Accumulation-layer states on bare InAs(110)

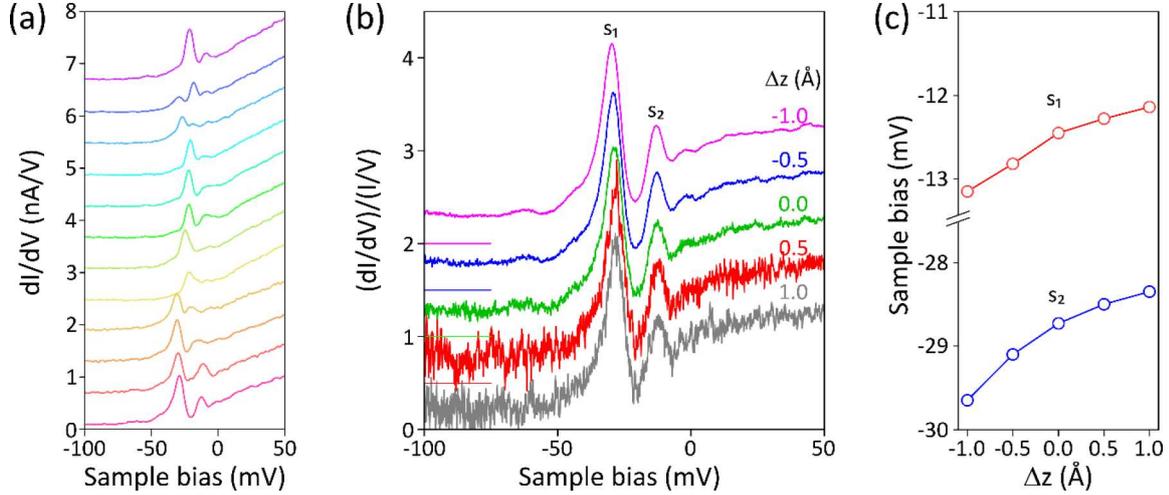

Fig. S2. (a) Conductance spectra taken at various lateral tip positions on the bare InAs(110) surface within an area of 3000 Å× 3000 Å in size. The accumulation-layer states give rise to conductance peaks right below the Fermi level at V=0, typically (but not always) appearing as a double peak; the spatial variation in peak magnitude and energy position is attributed to the effect of electrostatic disorder due to residual defects in the surface-near region. (b) Spectra showing the two conductance peaks associated with the first ($s_1$) and second subband ($s_2$) of the surface-accumulated electrons taken at the same lateral position but at varying tip height Δz (relative to the initial set-point tip height at 0.1 nA and 0.1 V). The normalized conductance (dI/dV)/(I/V) is plotted for a direct comparison between the spectra recorded at different tip-surface separation. The peaks (states) appear to be largely unaffected by the STM tip, only a small peak shift to lower energies is found at increasing tip proximity as shown in panel (c); the minor shift is attributed to the effect of the tip-induced electric field.

**S4. Vertical and lateral atom manipulation of In on InAs(110)**

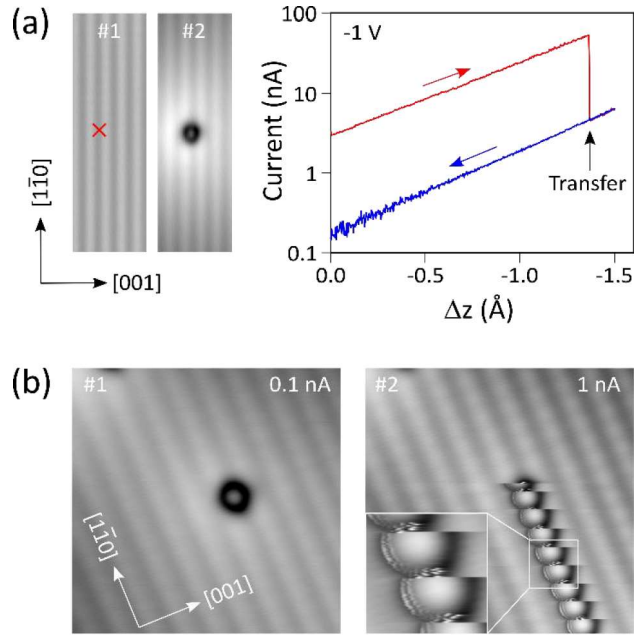

Fig. S4. (a) Left panel: STM images (0.1 nA, 0.1 V, 28 Å × 92 Å) of a surface area (#1) prior and (#2) after transferring an In atom from the tip apex to the surface at the position marked by the red cross (vertical atom manipulation). Right panel: tunnel current versus change in tip height Δz upon moving the tip towards the surface (red curve) and retracting it back to the initial separation defined by the set point parameters 0.1 nA and 0.1 V (blue curve); the atom drops from the tip to the surface at Δz=-1.38 Å. (b) A single In adatom (#1) imaged at 0.1 nA and 0.1 V followed by (#2) a scan (line-by-line from top to bottom) of the same surface area but at 1 nA tunnel current. As the scanning tip reaches the adatom position, it pulls the adatom along the channel via attractive force interaction (lateral atom manipulation). During this process, the adatom hops from one stable configuration to the other which obviously corresponds to the $I_{i2}$ site in the potential-energy surface shown in Fig. 1(b) of the main text.